\documentclass[aps,epsfig,floats,twocolumn,amssymb,amsmath,showpacs]{revtex4}
\usepackage{graphicx}
\begin{document}
\title{Brushes of flexible, semiflexible and rodlike diblock polyampholytes:
 Molecular dynamics simulation and scaling analysis}

\author{Majid Baratlo}
\affiliation{Institute for Advanced
Studies in Basic Sciences, Zanjan 45195-1159, Iran}

\author{Hossein Fazli}
\email{fazli@iasbs.ac.ir} \affiliation{Institute for Advanced
Studies in Basic Sciences, Zanjan 45195-1159, Iran}

\date{\today}

\begin{abstract}

Planar brushes of flexible, semiflexible and rodlike diblock
polyampholytes are studied using molecular dynamics simulations
in a wide range of the grafting density. Simulations show linear
dependence of the average thickness on the grafting density in
all cases regardless of different flexibility of anchored chains
and the brushes different equilibrium conformations. Slopes of
fitted lines to the average thickness of the brushes of
semiflexible and rodlike polyampholytes versus the grafting
density are approximately the same and differ considerably from
that of the brushes of flexible chains. The average thickness of
the brush of flexible diblock polyampholytes as a function of the
grafting density is also obtained using a simple scaling analysis
which is in good agreement with our simulations.

\end{abstract}
 \maketitle

\section{Introduction}  \label{intro}

Macromolecules containing ionizable groups when dissolve in a
polar solvent such as water, dissociate into charged
macromolecules and counterions (ions of opposite charge).
Depending on acidic or basic property of their monomers, ionizable
polymers in solution can be classified into polyelectrolytes and
polyampholytes. Polyelectrolytes contain a single sign of charged
monomers and polyampholytes bear charged monomers of both signs.
These macromolecules are often water-soluble and have numerous
industrial and medical applications. Many biological
macromolecules such as DNA, RNA, and proteins are charged
polymers. In polymer science, charged polymers has been an
important subject during last several decades
\cite{Forster,Barrat,rev_Dobrynin,Ruhe}. Contrary to a
polyelectrolyte chain in which the intra-chain electrostatic
interactions are repulsive and tend to swell the chain, in a
polyampholyte chain attractive interactions between charged
monomers of opposite sign tend to decrease the chain size.
Oppositely charged monomers can be distributed randomly along a
polyampholyte chain or charges of one sign can be arranged in long
blocks. With the same ratio of positively and negatively charged
monomers (isoelectric condition), behaviors of a single
polyampholyte and the solution of polyampholytes depend
noticeably on the sequence of charged monomers on the chains. For
example, it has been shown that the sequence of charged amino
acids (charge distribution) along ionically complementary
peptides affect the aggregation behavior and self-assembling
process in the solution of such peptides \cite{Hong,Jun}. Also,
using Monte Carlo simulations it has been shown that charged
monomers sequence of charge-symmetric polyampholytes affect their
adsorption properties to a charged surface \cite{Messina}.

The properties of the system of polymers anchored on a surface are
of great interest both in industrial and biological applications
and academic research. With a sufficiently strong repulsion
between the polymers, the chains become stretched and the
structure obtained is known as a polymer brush. Planar and curved
brushes formed by grafted homopolymers have extensively been
investigated by various theoretical methods
\cite{Wijmans,Lindberg,Klos,Almusallam}.
\begin{figure*}
\includegraphics[width=1.9 \columnwidth]{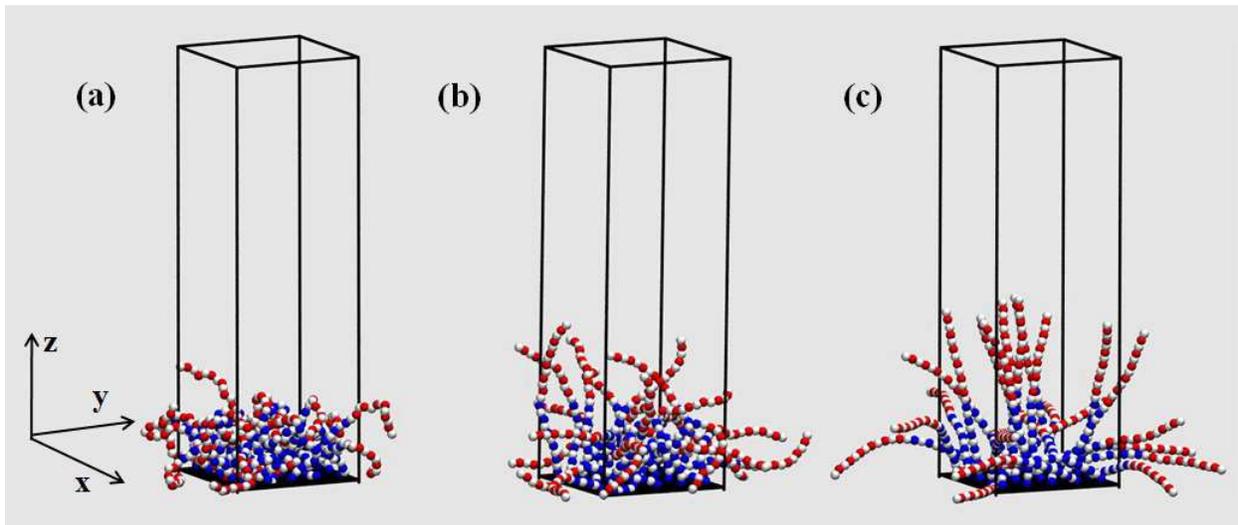}
\caption{(Color online) Sample snapshots of brushes of a)
flexible ($l_p=5\sigma$), b) semiflexible ($l_p=25\sigma$) and c)
rodlike ($l_p=200\sigma$) diblock polyampholytes at grafting
density $\rho_a\sigma^2=0.08$. Neutral , positively and negatively
charged monomers are shown by white, red (dark) and blue (black)
spheres respectively. Counterions are not shown and periodic
boundary condition is removed for clarity.} \label{fig1}
\end{figure*}
The anchored polymers of a brush may be consisting of charged
monomers. In this case, the brush is known as a polyelectrolyte
or a polyampholyte brush depending on the charged monomers of the
chains being of the same sign or being composed of both signs. In
a brush of charged polymers, electrostatic interactions introduce
additional length scales such as Bjerrum length and Debye
screening length to the system.

In a polyelectrolyte brush, the repulsion of electrostatic origin
between the chains can be sufficiently strong even at low
grafting densities, making it easy for the system to access the
brush regime. Polyelectrolyte brushes have been investigated
extensively using both theoretical
\cite{Pincus,Zhulina1,Zhulina2,Borisov2,Zhulina3,Naji1,Ahrens,Naji2}
and computer simulation methods
\cite{Naji1,Ahrens,Naji2,Csajka,Seidel,Fazli}. At high enough
grafting densities and charge fractions of polyelectrolyte chains,
most of counterions are trapped inside the polyelectrolyte brush
and competition between osmotic pressure of the counterions and
elasticity of the chains determines the brush thickness. This
regime of a polyelectrolyte brush is known as the osmotic regime
in which some theoretical scaling methods predict no dependence of
the brush thickness to the grafting density
\cite{Pincus,Borisov1}. However, other  scaling method that takes
into account the excluded volume effects and nonlinear elasticity
of polyelectrolyte chains and is in agreement with experiment and
simulation, predicts a linear dependence of the brush thickness on
the grafting density \cite{Naji1,Ahrens,Naji2}. Also, it has been
shown that diffusion of a fraction of counterions outside the
polyelectrolyte brush leads to a logarithmic dependence of the
average brush thickness on the grafting density \cite{Zhulina3}.

Electrostatic interactions in a polyelectrolyte brush cause most
of the counterions to be trapped inside the brush and help the
chains to be more stretched and the brush to be more aligned.
However, in a brush of overally neutral polyampholyte chains,
most of counterions are outside the brush and the electrostatic
correlations tend to decrease the chains size and the brush
thickness. At a given value of the grafting density, the average
thickness and equilibrium properties of such a polyampholyte
brush are mainly determined by the chains properties such as
fraction and sequence of charged monomers and the bending energy.
Brushes formed by grafted diblock polyampholytes have been
investigated by lattice mean field modeling
\cite{Shusharina2,Shusharina3} and computer simulation
\cite{Akinchina1}. The effect of chain stiffness, charge density
and grafting density on spherical brushes of diblock
polyampholytes and interaction between colloids with grafted
diblock polyampholytes have been studied using Monte Carlo
simulations \cite{Akinchina2,Linse}. Also, using molecular
dynamics (MD) simulations, the effects of various parameters such
as charged monomers sequence, grafting density and salt
concentration on the average thickness and equilibrium
conformations of planar semiflexible polyampholyte brushes have
been investigated \cite{Baratlo}.

In this paper, we study planar brushes of flexible, semiflexible
and rodlike diblock polyampholytes using MD simulations in a wide
range of the grafting density. We find that in all cases the
average brush thickness linearly depends on the grafting density
regardless of the chains different flexibility. Our results also
show that the strength of this dependence is considerably weaker
in the case of the brush of flexible polyampholytes than two
other cases. Despite mentioned same functionality obtained for
the average thickness versus the grafting density for brushes of
different polyampholytes we find that histograms of their
equilibrium conformations are noticeably different. Inter-chain
correlations are too weak in the brush of flexible polyampholytes
and the brush properties are dominantly determined by single
chain behavior. In this case, dependence of the equilibrium
conformation of the brush on the grafting density is very weak.
In the cases of the brushes of semiflexible and rodlike
polyampholytes however, because of the combination of
electrostatic correlations and strong excluded volume effects,
collective behavior of the chains is dominant and dependence of
equilibrium conformations on the grafting density is strong. In
these cases, we also observe separation of the anchored chains
into two coexisting fractions. Using a simple scaling method
which is consistently applicable for the brush of flexible chains,
we describe theoretically the linear dependence of the brush
thickness on the grafting density.

The rest of the paper is organized as follows. In Section
\ref{simulation} we describe our model and simulation method in
detail and present the results of MD simulations. Our scaling
analysis to describe dependence of the average thickness of
flexible diblock polyampholyte brushes on the grafting density is
presented in section \ref{theory}. In section \ref{discussion} we
conclude the paper and present a short discussion.

\section{Molecular dynamics simulation of diblock polyampholyte brushes}    \label{simulation}
\subsection{The model and the simulation details}
In our simulations which are performed with the MD simulation
package ESPResSo \cite{ESPResSo}, each brush is modeled by $M=25$
diblock polyampholyte bead-spring chains of length $N=24$ (24
spherical monomers) which are end-grafted onto an uncharged
surface at $z=0$. The positions of anchored monomers which are
fixed during the simulation, form an square lattice on the
grafting surface ($x-y$ plane) with lattice spacing
$d=\rho_{a}^{-1/2}$ in which $\rho_a$ is the grafting density of
the chains. The fraction $f=\frac{1}{2}$ of the monomers of each
chain are charged and the chains consist of an alternating
sequence of charged and neutral monomers. Each chain contains the
same number of positively and negatively charged monomers with
charges $e$ and $-e$ respectively (see Fig. \ref{fig1}). Excluded
volume interaction between particles is modeled by a shifted
Lennard-Jones potential,
\begin{equation}
 u_{LJ}(r) = \left\lbrace
  \begin{array}{l l}
    4\varepsilon\
    \{(\frac{\sigma}{r})^{12}-(\frac{\sigma}{r})^{6}+\frac{1}{4}\} & \text{if $r<r_{c}$},\\
    0 & \text{if $r \geq r_{c}$},
  \end{array}
\right. \label{slj}
\end{equation}
in which $\epsilon$ and $\sigma$ are the usual Lennard-Jones
parameters and the cutoff radius is $r_{c}=2^{1/6} \sigma$.
Successive monomers of each chain are bonded to each other by a
FENE (finite extensible nonlinear elastic) potential \cite{Grest},
\begin{equation}
u_{bond}(r)=\left\lbrace
     \begin{array}{l l}
       - \frac{1}{2}k_{bond}R_{0}^{2}\ln(1-(\frac{r}{R_{0}})^{2})& \text{if $r<R_{0}$},\\
       0 & \text{if $r \geq R_{0}$},
  \end{array}
\right. \label{FENE}
\end{equation}
with bond strength $k_{bond}=30\varepsilon/\sigma^{2}$ and maximum
bond length $R_{0}= 1.5\sigma$. Bending elasticity of the chains
is modeled by a bond angle potential,
\begin{equation}
u_{bend}(r)=k_{bend}(1-\cos\theta) \label{angle},
\end{equation}
in which $\theta$ is the angle between two successive bond vectors
and $k_{bend}$ is the bending energy of the chains. The value of
the persistence length, $l_p$, of the chains depends on the value
of $k_{bend}$ as $l_p=\frac{k_{bend}}{k_BT}\sigma$. To model
brushes of flexible, semiflexible and rodlike chains, we use four
different values of $k_{bend}$ namely $0$, $5k_BT$, $25k_BT$ and
$200k_BT$ respectively. The simulation box is of volume $L \times
L \times L_z$ in which $L$ is the box width in $x$ and $y$
directions and $L_z$ is its height in $z$ direction and the
grafting density is given by $\rho_{a}=M/L^{2}$. We consider
$M\times N\times f$ monovalent counterions to neutralize the
chains charge. Positive and negative monovalent counterions are
modeled by equal number of spherical Lennard-Jones particles of
diameter $\sigma$ with charges $e$ and $-e$ respectively. All the
particles interact repulsively with the grafting surface at short
distances with the shifted Lennard-Jones potential introduced in
Eq. \ref{slj}. In addition, a similar repulsive potential is
applied at the top boundary of the simulation box and in our
simulations $L_z =2N\sigma$. All the charged particles interact
with each other with the Coulomb interaction
\begin{equation}
u_{C}(r)= k_{B}Tq_{i}q_{j}\frac{l_{B}}{r} \label{coul}
\end{equation}
in which $q_{i}$ and $q_{j}$ are charges of particles $i$ and $j$
in units of elementary charge $e$ and $r$ is separation between
them. The Bjerrum length, $l_B$, which determines the strength of
the Coulomb interaction relative to the thermal energy, $k_BT$, is
given by $l_{B}=e^{2}/\varepsilon k_{B}T$, where $\varepsilon$ is
the dielectric constant of the solvent and we set $l_{B}=2\sigma$
in our simulations. Periodic boundary conditions are applied only
in two dimensions ($x$ and $y$). To calculate Coulomb forces and
energies, we use the so-called $MMM$ technique introduced by
Strebel and Sperb \cite{Strebel} and modified for laterally
periodic systems ($MMM2D$) by Arnold and Holm \cite{Arnold}. The
temperature in our simulations is kept fixed at
$k_{B}T=1.2\varepsilon$ using a Langevin thermostat.
\begin{figure}[t]
\includegraphics[width=1.0 \columnwidth]{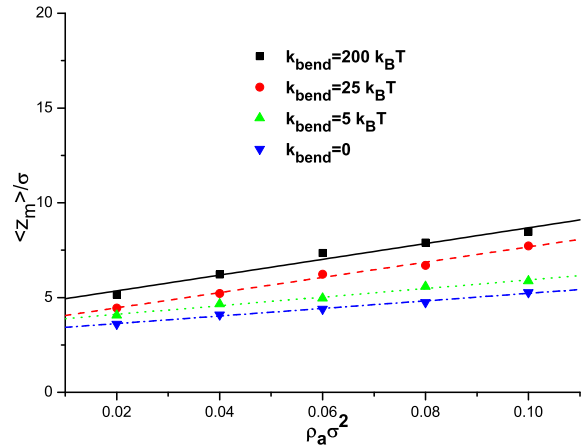}
\caption{(Color online) The average thickness of the brushes
formed by flexible ($k_{bend}=0$ and $5k_BT$), semiflexible
($k_{bend}=25k_BT$) and rodlike ($k_{bend}=200k_BT$) diblock
polyampholytes versus dimensionless grafting density,
$\rho_{a}\sigma^{2}$. The lines are linear fits to our simulation
data. The slopes of solid, dashed, dotted and dash-dotted lines
are 41.6, 40.2 and 22.7 and 20.0 respectively. The size of the
symbols corresponds to the size of error bars.} \label{fig2}
\end{figure}

For each value of the bending energy, $k_{bend}$, we do
simulations of the brush at dimensionless grafting densities
$\rho_{a}\sigma^{2}=0.02,0.04,0.06,0.08,0.10$. In the beginning of
each simulation, all of the chains are straight and perpendicular
to the grafting surface and all the ions are randomly distributed
inside the simulation box. We equilibrate the system for
$1.6\times 10^6$ MD time steps which is enough for all values of
the grafting density mentioned above and then calculate thermal
averages over $1500$ independent configurations of the system
selected from $2.25\times 10^6$ additional MD steps after
equilibration. MD time step in our simulations is
$\tau=0.01\tau_0$ in which
$\tau_0=\sqrt{\frac{m\sigma^2}{\varepsilon}}$ is the MD time scale
and $m$ is the mass of the particles.

We calculate the average brush thickness which can be measured by
taking the first moment of the monomer density profile
\begin{equation}
\langle z_{m} \rangle =\frac{\int_{0}^{\infty}{z
\rho_{m}(z)}{dz}}{\int_{0}^{\infty}{ \rho_{m}(z)}{dz}},\label{z_m}
\end{equation}
in which $\rho_{m}(z)$ is the number density of monomers as a
function of the distance from the grafting surface. For a better
monitoring of the statistics of the chains conformations, we
calculate the histogram of the mean end-to-end distance of the
chains, $P(R)$, in which
$R=\frac{1}{M}\sum_{i=1}^{M}|\textbf{R}_i|$ and $\textbf{R}_i$ is
the end-to-end vector of chain $i$. We also calculate the
histogram of the average distance of the end monomers of the
chains from the grafting surface, $P(z_{end})$, in which
$z_{end}=\frac{1}{M}\sum_{i=1}^{M}z_i$ and $z_i$ is the $z$
component of the end monomer of chain $i$. With the same method
discussed in Ref. \cite{Baratlo}, it has been checked that our
results are not affected by finite-size effects (see Sec.
\ref{discussion}).

\subsection{Results}

\begin{figure}[t]
\includegraphics[width=0.90 \columnwidth]{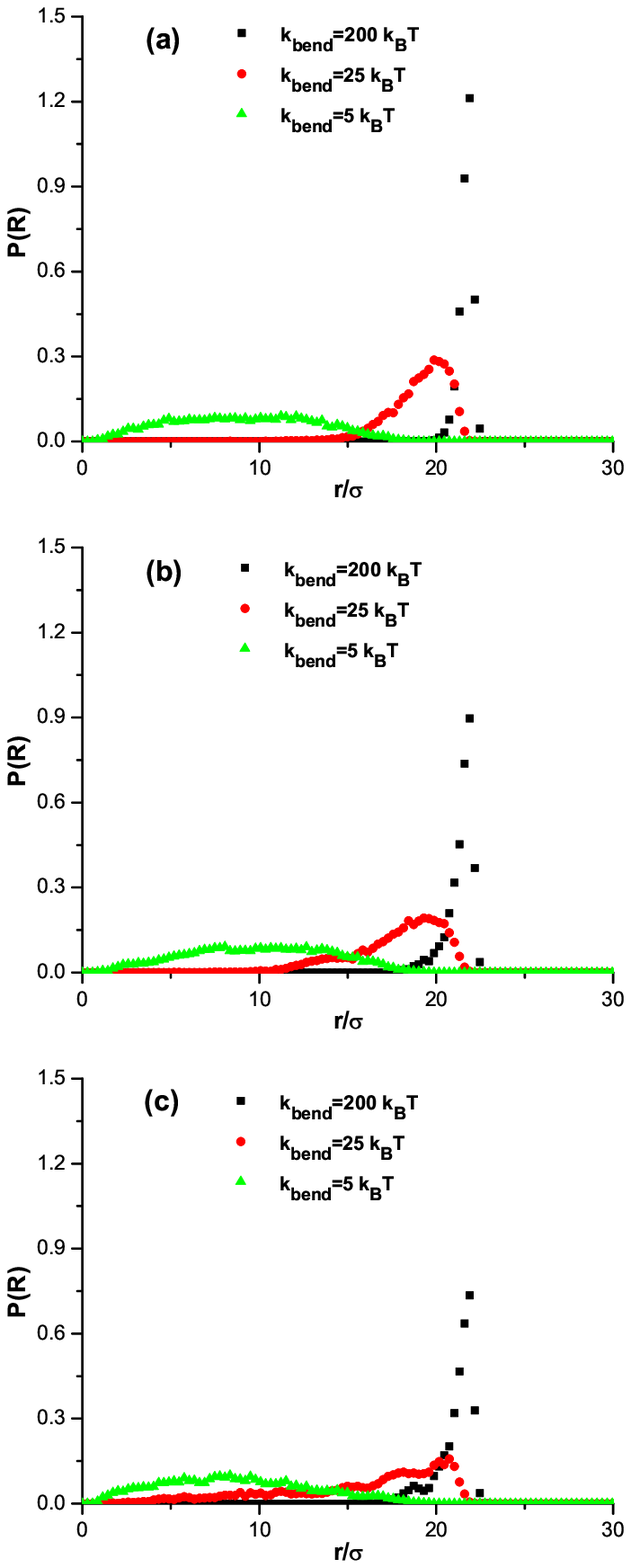}
\caption{(Color online) Histogram of the average end-to-end
distance of the chains for brushes formed by flexible,
semiflexible and rodlike diblock polyampholytes at grafting
densities a) $\rho_{a}\sigma^{2}=0.02$, b)
$\rho_{a}\sigma^{2}=0.06$ and c) $\rho_{a}\sigma^{2}=0.10$.}
\label{fig3}
\end{figure}
\begin{figure}[t]
\includegraphics[width=0.90 \columnwidth]{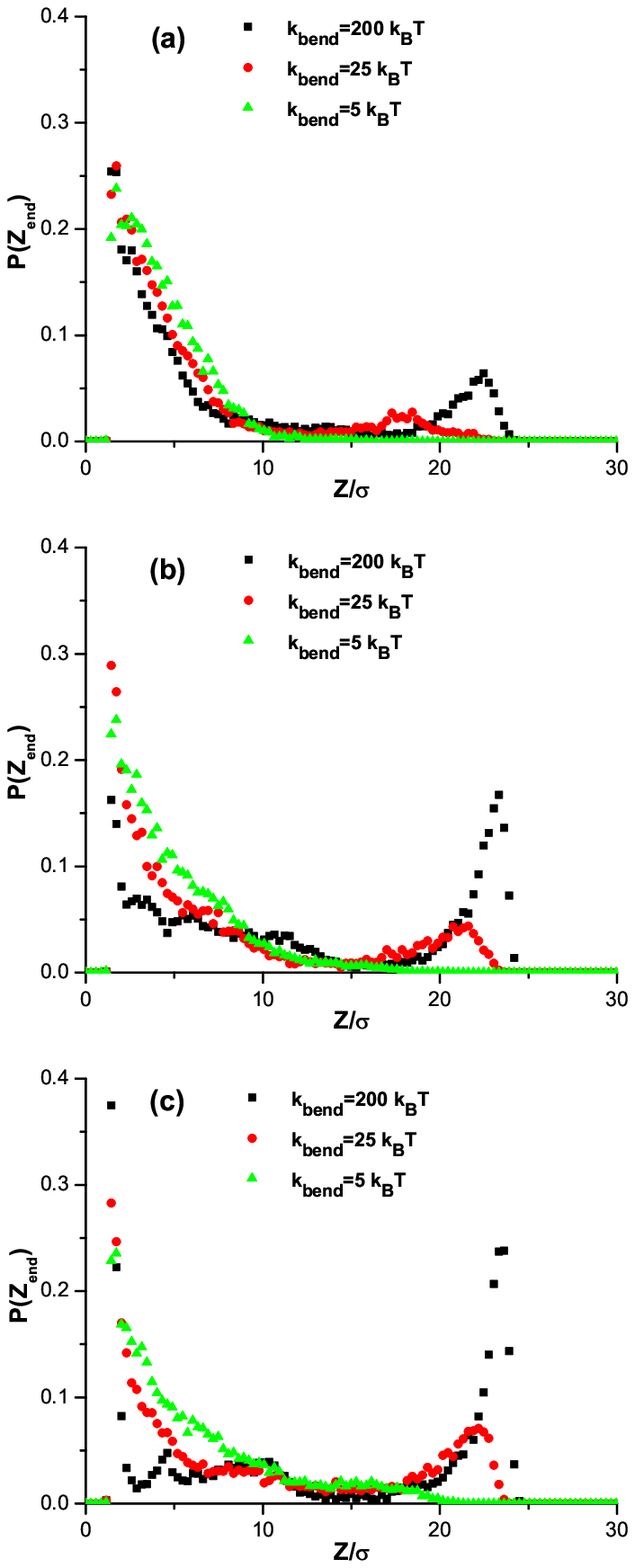}
\caption{(Color online) Histogram of the average distance of the
end monomers of the chains from the grafting surface for the
brushes of flexible, semiflexible and rodlike diblock
polyampholytes at grafting densities a)
$\rho_{a}\sigma^{2}=0.02$, b) $\rho_{a}\sigma^{2}=0.06$ and c)
$\rho_{a}\sigma^{2}=0.10$.} \label{fig4}
\end{figure}

The average thickness versus the grafting density for brushes of
flexible, semiflexible and rodlike diblock polyampholytes are
shown in Fig. \ref{fig2}. Dependence of the average thickness on
the grafting density can be described well by a linear function
for all values of the bending energy, $k_{bend}$, that we use in
our simulations. Also, it can be seen that at all values of the
grafting density the average brush thickness decreases with
increasing the flexibility of the chains. Linear fits to the
average thickness of the brushes of semiflexible and rodlike
chains for which the persistence length, $l_p$, exceeds their
contour length, $L_c$, are of approximately the same slope (see
solid and dashed lines in Fig. \ref{fig2}). The slopes of the
linear dependence in cases of two brushes of flexible chains are
also approximately the same and differ by a factor of $\sim
\frac{1}{2}$ from those of two other cases (dotted and
dash-dotted lines in Fig. \ref{fig2}). To analyze such
dependencies of the brushes thickness on the grafting density, we
look at the equilibrium conformations statistics of the chains at
different values of $\rho_a$. In Figs. \ref{fig3} and \ref{fig4}
the histograms $P(R)$ and $P(z_{end})$ at three different grafting
densities are shown for brushes of flexible, semiflexible and
rodlike chains. Because of very similar behaviors of the
histograms in $k_{bend}=0$ and $k_{bend}=5k_BT$ cases, the
histograms of the brush with $k_{bend}=0$ are not shown for
clarity of the figures. As it can be seen in Fig. \ref{fig3}, in
the case of the brush of flexible diblock polyampholytes
($k_{bend}= 5k_BT$), dependence of the histogram profile on the
grafting density is very weak. Also, in this case, the
contribution of large values of $r$ ($r\simeq L_c$) in the
histogram is negligible which shows that the chains are mostly
coiled at all grafting densities (see a sample configuration of
the brush at $\rho_a\sigma^2=0.08$ in Fig. \ref{fig1}a). In the
case of the brush of semiflexible chains ($k_{bend}= 25k_BT$),
with increasing the grafting density, the values of the histogram
corresponding to smaller values of $r$ become nonzero showing
that polyampholyte chains take buckled conformations at high
grafting densities \cite{Baratlo}. As it is expected, the
histogram $P(R)$ of the brush of rodlike chains ($k_{bend}=
200k_BT$) exhibit no noticeable buckling of the chains. The
histograms $P(z_{end})$ in Fig. \ref{fig4} show that in the brush
of flexible chains at all grafting densities the end monomers are
mostly distributed near the grafting surface showing that the
positive blocks of the chains are mostly turned back towards the
anchored negative blocks. The profile of this histogram also
doesn't depend noticeably on the grafting density. The histograms
$P(z_{end})$ for brushes of semiflexible and rodlike chains show
that two maxima appear and their height increase with increasing
the grafting density. By combining the information obtained from
histograms $P(R)$ and $P(z_{end})$ for the brush of rodlike
chains it can be understood that at high grafting densities a
fraction of the chains which are perpendicular to the grafting
surface coexist with the remaining fraction which fluctuate in
the vicinity of the grafting surface. In the case of the brush of
semiflexible chains also, the fraction of perpendicular chains to
the brush surface coexist with the chains which are buckled
towards the grafting surface \cite{Baratlo}. Histograms $P(R)$
and $P(z_{end})$ show that conformations of the brushes of
semiflexible and rodlike chains change noticeably with changing
the grafting density despite the case of the brush of flexible
chains.

\section{Linear dependence of the brush thickness on the grafting density: Scaling approach }    \label{theory}

As mentioned in the previous section, our MD simulations show that
the brush thickness is a linear function of the grafting density
regardless of different values of grafted diblock polyampholytes
bending energy. Also, as it is shown in Fig. \ref{fig2}, in cases
of the brushes of flexible chains the slope of the fitted line to
the average thickness versus the grafting density is considerably
smaller than that in two other cases. We use here a simple
scaling theory similar to that of the solution of
charge-symmetric diblock polyampholytes \cite{Shusharina1} to
describe the linear dependence of the brush thickness on the
grafting density. This scaling analysis is applicable to the
brushes of flexible diblock polyampholytes.
\begin{figure}[t]
\includegraphics[width=.99 \columnwidth]{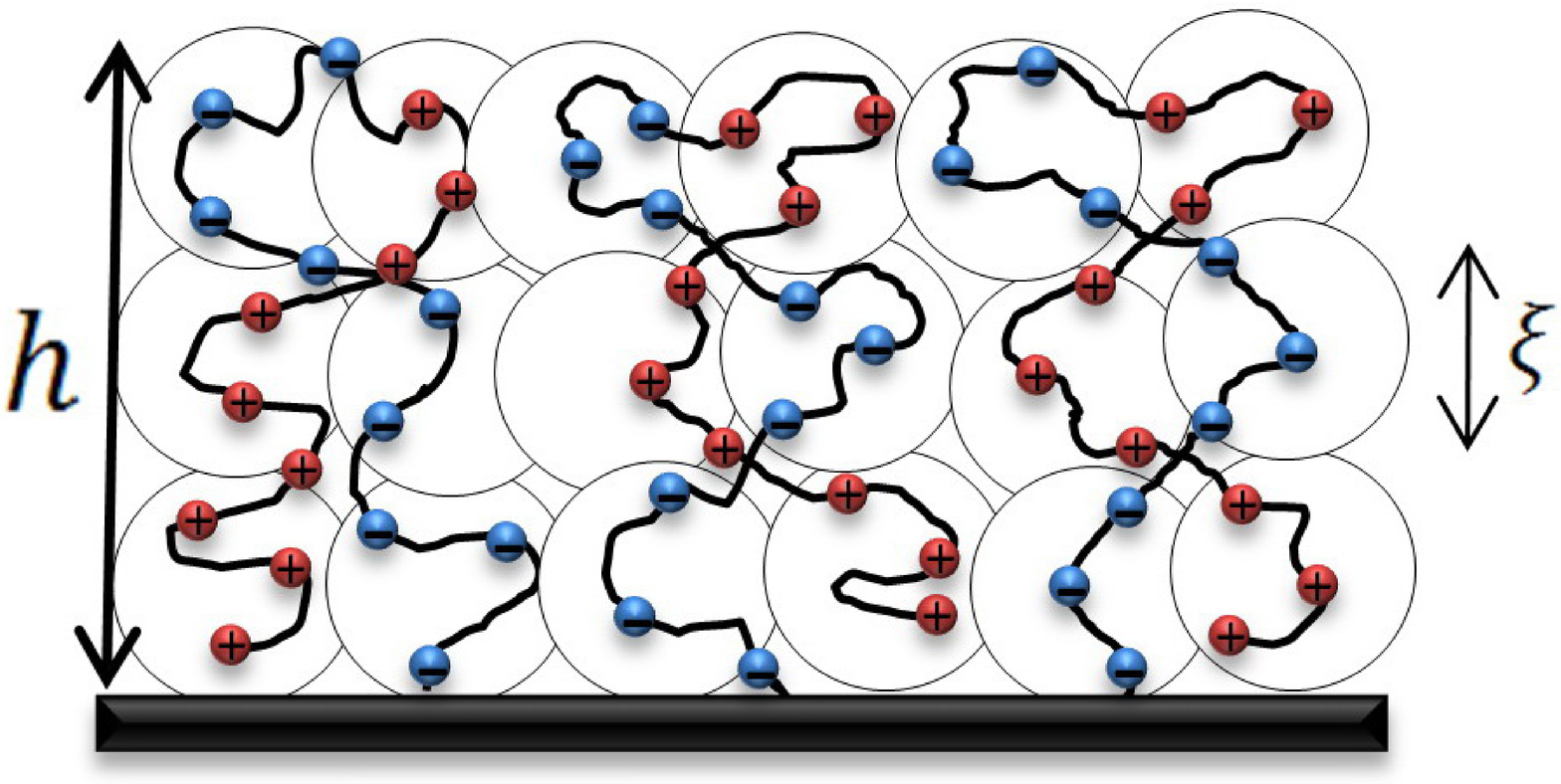}
\caption{(Color online) The blob picture of a flexible diblock
polyampholyte brush. The blobs of positive charge are more
probably surrounded by blobs of negative charge. Positively and
negatively charged monomers are shown by red ($+$) and blue ($-$)
sphere respectively. The neutral monomers are not shown for
clarity. $h$ and $\xi$ are the brush thickness and correlation
blob size respectively.} \label{fig5}
\end{figure}

Consider $M$ flexible diblock polyampholyte chains end-grafted to
a flat surface of area $A$. The degree of polymerization of each
chain and the fraction of charged monomers are $N$ and $f$
respectively. Electrostatic attractive interactions between
oppositely charged blocks of the chains lead them to form a dense
layer of positive and negative monomers of average thickness $h$
near the grafting surface (see Fig. \ref{fig5}). Inside the
layer, to use the blob concept we define correlation length $\xi$
as a length scale that electrostatic interactions don't perturb
the chains statistics at smaller length scales and are dominant
over thermal fluctuations at larger length scales. Let suppose
that the number of monomers inside the correlation blob is $g$
and we have $\xi\approx b g^{\nu}$ where $b$ and $\nu$ are the
Kuhn length and the Flory exponent respectively. Accordingly, the
layer is a melt of correlation blobs in which positive blobs with
a high probability are surrounded by negative blobs.
Electrostatic interaction between any two neighboring blobs is of
the order of the thermal energy, $k_BT$
\begin{equation}
k_{B}T \mid\frac{l_{B} f^{2} g^{2}}{\xi}\mid\approx k_{B}T
\label{elec-kt}.
\end{equation}
Thus we obtain $\xi \approx l_{B}f^{2}g^{2}\approx b (\frac{l_{B}
f^{2}}{b})^{\frac{\nu}{\nu-2}}$. As a result, the local monomer
concentration inside the blob is
\begin{equation}
\rho_{local}\approx \frac{g}{\xi^{3}}\approx
\frac{\xi^{\frac{1-3\nu}{\nu}}}{b^{\frac{1}{\nu}}}\approx
\frac{1}{b^{3}}(\frac{l_{B}f^{2}}{b})^{\frac{1-3\nu}{\nu-2}}.
\label{den-local}
\end{equation}
Considering that correlation blobs are space filling, the global
monomer concentration, $\rho_{global}=\frac{M N}{A
h}=\rho_{a}\frac{N}{h}$, approximately equals to the local
monomer concentration, $\rho_{global}\approx \rho_{local}$. Thus,
dependence of the thickness of a brush of diblock polyampholytes
on the grafting density can be obtained as
\begin{equation}
h\approx\rho_{a} N b^{3} (\frac{l_{B}}{b}
f^{2})^{\frac{3\nu-1}{\nu-2}} \label{thickness-1}.
\end{equation}
The main prediction of this equation in the range of its validity
is as follows. The brush thickness, $h$, is a linear function of
the grafting density, $\rho_a$, irrespective of the values of the
system parameters. Only the slope of this linear function,
$\alpha=N b^{3} (\frac{l_{B}}{b} f^{2})^{\frac{3\nu-1}{\nu-2}}$,
depends on the values of the system parameters.

One should note here that despite the bulk solution of flexible
diblock polyampholytes, in a polyampholyte brush the fact that
the chains are end grafted to the surface introduces an
additional length scale to the system, namely $d=\rho_a^{-1/2}$.
In the scaling analysis presented here, if the size of the
correlation length, $\xi$, exceeds the separation between the
grafting points, $d$, the scaling approach becomes inconsistent.
Although linear dependence of the brush thickness on the grafting
density is observed at all values of the bending energy used in
our simulations, the condition $\xi<d$ is valid only in the case
of the brush of flexible chains with $k_{bend}=0$. In this case
the values of $\frac{\xi}{d}$ corresponding to lowest and highest
values of the grafting density we have used in our simulations
are 0.19 and 0.45 respectively. For the set of parameters used in
our simulations, in the case of the brush with $k_{bend}=0$ the
upper limit of the validity of the scaling method ($\xi\simeq d$)
corresponds to the grafting density $\rho_a\sigma^2\simeq 0.55$
which is quite far from our range of the grafting density. For a
brush of flexible chains with $k_{bend}=0$ using the values
$b\simeq\sigma$ and $\nu\simeq\frac{3}{5}$ gives the slope
$\alpha_f\simeq 35$ for the brush thickness versus the grafting
density for the used set of system parameters. As it is shown in
Fig. \ref{fig2}, the value of $\alpha_f$ obtained from our
simulations of flexible chains is $\alpha_f\simeq20.0$ which is
in reasonable agreement with prediction of the scaling method.

Our scaling method is not consistently applicable for the brush of
flexible chains with $l_p=5\sigma$ because the range of the
grafting density used in our simulations is higher than the
validity range of the condition $\xi<d$ in this case. This method
is not also applicable for the brushes of semiflexible and
rodlike chains because in these cases the Kuhn length exceeds the
contour length of the chains. The fact that the average thickness
linearly depends on the grafting density at all values of
$k_{bend}$ used in our simulations shows that although the
scaling method presented here can not be used to describe all the
simulation results, this linear dependence persists over a wide
range of the system parameters.

\section{conclusions and discussion} \label{discussion}

Brushes of flexible, semiflexible and rodlike diblock
polyampholytes have been studied using MD simulations and a
scaling analysis has been presented to describe the results of
the simulation of flexible chins brush. The average thickness as a
function of the grafting density and histograms of equilibrium
conformations of the brushes are obtained. Strong dependence of
the system conformations on the grafting density and separation
of the chains into two coexisting fractions at high grafting
densities have been observed in cases of the brushes of
semiflexible and rodlike chains. In cases of the brushes of
flexible chains, single-chain behavior is dominant and dependence
of the brush conformations on the grafting density is very weak.
In spite of above mentioned differences, it has been observed that
dependence of the average brush thickness on the grafting density
is linear for brushes of all different chains. This linear
dependence resulted from our MD simulations has been described
well using a simple scaling method in the case of the brush of
flexible chains.

Brushes of polyelectrolytes and polyampholytes are dense assembly
of these macromolecules in which the interplay between
electrostatic correlations, strong excluded volume effects and
bending elasticity of the chains determine equilibrium properties
of the system. The main differences between brushes of
polyelectrolyte and polyampholyte chains originates from opposite
trends of inter- and intra-chain electrostatic interactions and
different rules of counterions osmotic pressure in these brushes.
In a brush of polyelectrolyte chains most of counterions are
contained inside the brush and their osmotic pressure tends to
increase the brush thickness. In a brush of overally neutral
polyampholytes however counterions are outside the brush and have
no effect on the brush thickness. Linear dependence of
polyelectrolyte brush thickness on the grafting density has
theoretically been described \cite{Naji1,Ahrens,Naji2}. The
results of a recent simulation of semiflexible polyampholytes
\cite{Baratlo} and our simulations and theoretical analysis here
show that linear dependence of the average thickness on the
grafting density is also the case in the brush of polyampholyte
chains. Amount of the flexibility of the chains which control the
strength of excluded volume effects and inter-chain correlations,
causes the equilibrium conformations of the brushes of three
different polyampholytes to be different. In brushes of
semiflexible and rodlike polyampholytes, strong dependence of the
equilibrium conformations on the grafting density and separation
of the chains into two coexisting fractions at high grafting
densities are resulted from strong electrostatic and excluded
volume correlations between the chains. Similar phenomenon has
been observed in the brush of rodlike polyelectrolytes
\cite{Fazli}.

To pay attention to possible finite-size effects in our
simulations, we have calculated the average size of the chains
lateral fluctuations as well as the histograms $P(z_{end})$ and
$P(R)$ for a larger brush containing $M=8\times8=64$ semiflexible
chains at the grafting density $\rho_a\sigma^2=0.1$
\cite{Baratlo}. The average size of lateral fluctuations is
defined as $\langle
\l_{lat}\rangle=\frac{1}{M}\sum_{i=1}^{M}\langle R_{i||}\rangle$,
where $R_{i||}=|\textbf{R}_i-\textbf{R}_i.\hat{z}|$ and
$\langle...\rangle$ denotes averaging over equilibrium
configurations. We have found that the histograms $P(R)$ and
$P(z_{end})$ do not change noticeably with increasing the system
size \cite{Baratlo}. Also, we have found that the value of
$\langle l_{lat}\rangle$ is smaller than the lateral size, $L$, of
the simulation box of a brush containing $M=25$ chains at the same
grafting density. This result shows that in simulation of the
brush of $M=25$ polyampholytes, the chains do not overlap with
their own images. Accordingly, to avoid time consuming
simulations of larger brushes, we concentrated on the simulations
of the brushes containing $M=25$ chains.

\begin{acknowledgments}

We would like to acknowledge A. Naji for his useful suggestions.

\end{acknowledgments}

\end{document}